\documentclass[aip,graphicx,amssymb,preprint]{revtex4-1}
\pdfoutput=1

\usepackage{siunitx}
\usepackage{hyperref}
\usepackage[utf8]{inputenc}
\usepackage[T1]{fontenc}
\usepackage{graphicx}
\usepackage{mathptmx}
\usepackage{dcolumn}
\usepackage{amsmath}
\usepackage{bm}

\usepackage{graphicx}

\newcommand{\angstrom}{\mbox{\normalfont\AA}}

\graphicspath{{Images/}}

\begin{document}
\title{Force Field Analysis Software and Tools (FFAST): Assessing Machine Learning Force Fields Under the Microscope}

\author{Gregory Fonseca}
\affiliation{Department of Physics and Materials Science, University of Luxembourg, L-1511 Luxembourg City, Luxembourg}
\author{Igor Poltavsky}
\affiliation{Department of Physics and Materials Science, University of Luxembourg, L-1511 Luxembourg City, Luxembourg}
\author{Alexandre Tkatchenko}
\email{alexandre.tkatchenko@uni.lu}
\affiliation{Department of Physics and Materials Science, University of Luxembourg, L-1511 Luxembourg City, Luxembourg}

\date{11 August 2023}

\begin{abstract}

As the sophistication of Machine Learning Force Fields (MLFF) increases to match the complexity of extended molecules and materials, so does the need for tools to properly analyze and assess the practical performance of MLFFs. To go beyond average error metrics and into a complete picture of a model's applicability and limitations, we develop FFAST (Force Field Analysis Software and Tools): a cross-platform software package designed to gain detailed insights into a model’s performance and limitations, complete with an easy-to-use graphical user interface. The software allows the user to gauge the performance of many popular state-of-the-art MLFF models on various popular dataset types, providing general prediction error overviews, outlier detection mechanisms, atom-projected errors, and more. It has a 3D visualizer to find and picture problematic configurations, atoms, or clusters in a large dataset. In this paper, the example of the MACE and Nequip models are used on two datasets of interest -- stachyose and docosahexaenoic acid (DHA) -- to illustrate the use cases of the software. With it, it was found that carbons and oxygens involved in or near glycosidic bonds inside the stachyose molecule present increased prediction errors. In addition, prediction errors on DHA rise as the molecule folds, especially for the carboxylic group at the edge of the molecule. We emphasize the need for a systematic assessment of MLFF models for ensuring their successful application to study the dynamics of molecules and materials.

\end{abstract}
\maketitle

\section{Introduction}

Many of the cornerstones of modern advancements in industry and technology rely on the introduction of novel or optimized materials, newly designed drugs, and the understanding of physico-chemical phenomena on a molecular level. Here, computer simulations and predictions of properties play a pivotal role in enabling, enhancing, or accelerating research and development.~\cite{leeComprehensiveReviewCurrent2019, ramosSheddingLightInteraction2019, mansbachSnailsSilicoReview2019, jingPolarizableForceFields2019, ekinsNextEraDeep2016, smalleyAIpoweredDrugDiscovery2017, eltonApplyingMachineLearning2018, eltonDeepLearningMolecular2019, faberMachineLearningEnergies2016, naserifarArtificialIntelligenceQM2021, covaDeepLearningDeep2019}

Amidst the computational tools, Machine Learning Force Fields (MLFF) are steadily rising in popularity, with a wide variety of models achieving remarkable predictive accuracy~\cite{nequip,  e3nn, mace1, mace2, GAP, SOAP, behlerPerspectiveMachineLearning2016, behlerGeneralizedNeuralNetworkRepresentation2007, schnet, schnet2, schnet3, schnetpack, ruppFastAccurateModeling2012, bartokMachineLearningUnifies2017, physnet, spookynet, unkeMachineLearningForce2021, sgdml, sgdml2, bigdml, orbnet, smithANI1ExtensibleNeural2017, devereuxExtendingApplicabilityANI2020, deringerRealisticAtomisticStructure2018, ryczkoConvolutionalNeuralNetworks2018, so3krates, christensenFCHLRevisitedFaster2020, wangMachineLearningCoarseGrained2019, koGeneralPurposeMachineLearning2021, kocerNeuralNetworkPotentials2022, gasteiger2022directional, gemnet, deringerGaussianProcessRegression2021, bartokMachineLearningGeneralPurpose2018, thomas2018tensor, khorshidiAmpModularApproach2016, liMolecularDynamicsOntheFly2015, keithCombiningMachineLearning2021, koFourthgenerationHighdimensionalNeural2021, yaoTensorMol0ModelChemistry2018, grisafiIncorporatingLongrangePhysics2019} for various molecules and materials of ever-increasing sizes and complexity~\cite{qm7x, qm7, qm9_1, qm9_2, gdml, md22, smithANI1ccxANI1xData2020}. The ultimate goal of MLFF is to bring the quality of expensive \textit{ab initio} methods to systems of larger scales, which is currently mostly feasible for efficient but significantly less accurate empirical mechanistic force fields. MLFFs aim to advance the FF development by learning highly accurate data generated by state-of-the-art quantum chemistry methods to then reproduce their results in a fraction of the computational time. However, in order to use these models in practice, it is crucial to understand which tasks a given MLFF is appropriate for. ML models are very sensitive to their training data, the training procedure, hyperparameters and many other details: two models with similar overall accuracy but trained on different data, hyperparameters, or architectures can still present fundamental differences in actual applications (for example molecular dynamics) that are not captured by simple error metrics~\cite{MLFF, forcesnotenough}.

For practical applications, it is important to know that our model is accurate and when it is likely to fail. This information is crucial, given that a single configuration that the model predicts inaccurately can move the system under study into an improbable or unphysical state, which could affect the remaining simulation. The probability of such a misprediction is steadily growing with increasing size, chemical, and structural complexity of the systems under study. While no silver bullets exist to make ML models perfectly stable, analyzing models beyond usual error metrics can avoid many pitfalls and limitations. This is why we developed FFAST (Force Field Analysis Software and Tools): a cross-platform software package designed to gain detailed insights into a model's performance and limitations, complete with an easy-to-use graphical user interface. Note that while this paper's main focus is MLFFs, all the described tools are equally applicable to empirical force fields or -- in fact -- force fields of any kind. 

The article is organized as follows: in the Software section, the software and its components are listed and briefly explained. In the Workflow section, part of those components are explored in a realistic use-case from start to finish. The Applications section applies the above workflow to actual datasets and models and analyses the results. Conclusions and outlooks can be found in the final section.

\section{Software}

Out of the box, FFAST can load a variety of MLFF models (currently supported sGDML~\cite{sgdml}, SchNet~\cite{schnet}, Nequip~\cite{nequip, e3nn}, MACE~\cite{mace1, mace2}, SpookyNet~\cite{spookynet}, and pre-predicted forces/energies) as well as dataset formats (currently supported .xyz, .npz, and .db). The ML models are used to generate \textbf{predictions of the energy and forces} on the loaded datasets. To prevent excessive loading times when handling large datasets or expensive models, this step can be accomplished in \textbf{headless mode} (i.e., without a graphical interface) to pre-compute the predictions externally (e.g., on a high-performance computer). 

Once loaded, various analysis tools are available in a user-friendly interface. 
\begin{itemize}
    \item \textbf{Error distributions} on both energy and force predictions are available, able to visualize multiple dataset/model combinations for easy comparisons. By default, mean average errors (MAE) are used, and the plots visualize a gaussian kernel-density estimate.
    \item  \textbf{Error timelines} show the MAE across the given time-ordered dataset (e.g., a dynamic). An adjustable smoothing factor is provided to disregard fluctuations by averaging each point over the given number of neighbors. 
    \item Plots of \textbf{cluster errors} provide a way to find problematic regions of configurational space. By default, agglomerative clustering on pairwise interatomic distances is followed up by KMeans on energies for a total of 40 clusters.
    \item \textbf{Error scatter} plots are available to find outliers quickly, both for energies and forces.
    \item Distribution of \textbf{atomic errors} can be used to determine the difference in prediction between chosen elements. The procedure is otherwise identical to error distributions.
\end{itemize}

The user can zoom in on most plots and choose to create \textbf{subdatasets}. These are subsets of loaded datasets to be further analyzed by the software. Subsets and datasets can be \textbf{saved in any format} that is provided by the software in case external editing or analysis is needed.

The program comes equipped with a \textbf{3D visualization tool}. The molecular structures inside of a dataset or subset can be viewed in one or more interactive windows. The visualizer comes with convenience features like \textbf{aligning geometries} along a chosen plane for easier visualization, \textbf{choosing the bonds} to be visualized, or \textbf{animating the molecule} throughout the selected dataset. One can extract information on the geometry such as \textbf{atomic distances, angles or dihedrals}. The user can enable the plotting of \textbf{force prediction errors on each atom} of the molecule, either at the given timeframe or averaged throughout the dataset/subset. Finally, \textbf{atoms of interest} can be selected and analyzed independently in all analysis tools above. 

One can also use the above-described tools to \textbf{analyze the underlying reference} dataset by creating a dummy model. In this case, FFAST will operate on the reference energy and forces instead of the difference between the reference data and ML models' prediction.

The software is designed to be modular, allowing users comfortable with Python to add features they need for their workflow. The workflow is streamlined such that swapping between software codes and separate scripts for different models is eliminated as much as possible. The only dependencies are those of the external ML models themselves (subject to their own installation process) and readily available Python packages. As such, FFAST provides a platform that beginners and experts alike can use to quickly and precisely assess their force field performance in an unbiased way.

\section{Workflow}

To illustrate the usefulness of the program, an example workflow is presented in this chapter. 
It is important to note that this use-case serves to guide and highlight important points of the software, not to be an all-inclusive exposition of all its features. 

As a first step, one or multiple ML models and datasets of interest can be loaded. If the datasets are large or the MLFFs are expensive, the program can be used in headless mode on a supercomputer to pre-compute the forces and energies. This allows the user to avoid unnecessary recalculations and prolonged loading times while using the software. For all applications in this paper, data for molecular systems are taken from the MD22 dataset~\cite{md22}. This chapter specifically focuses on its stachyose trajectory as an example. The particular selection of models used as illustrations in this paper are Nequip and MACE. The parameters used for all models reflect those recommended in the original paper or official software description of the respective packages. In all cases, all models are trained on the same 1000 training points for each dataset, respectively. 

\begin{figure}[hb!]
    \centering
    \includegraphics[width=.70\textwidth]{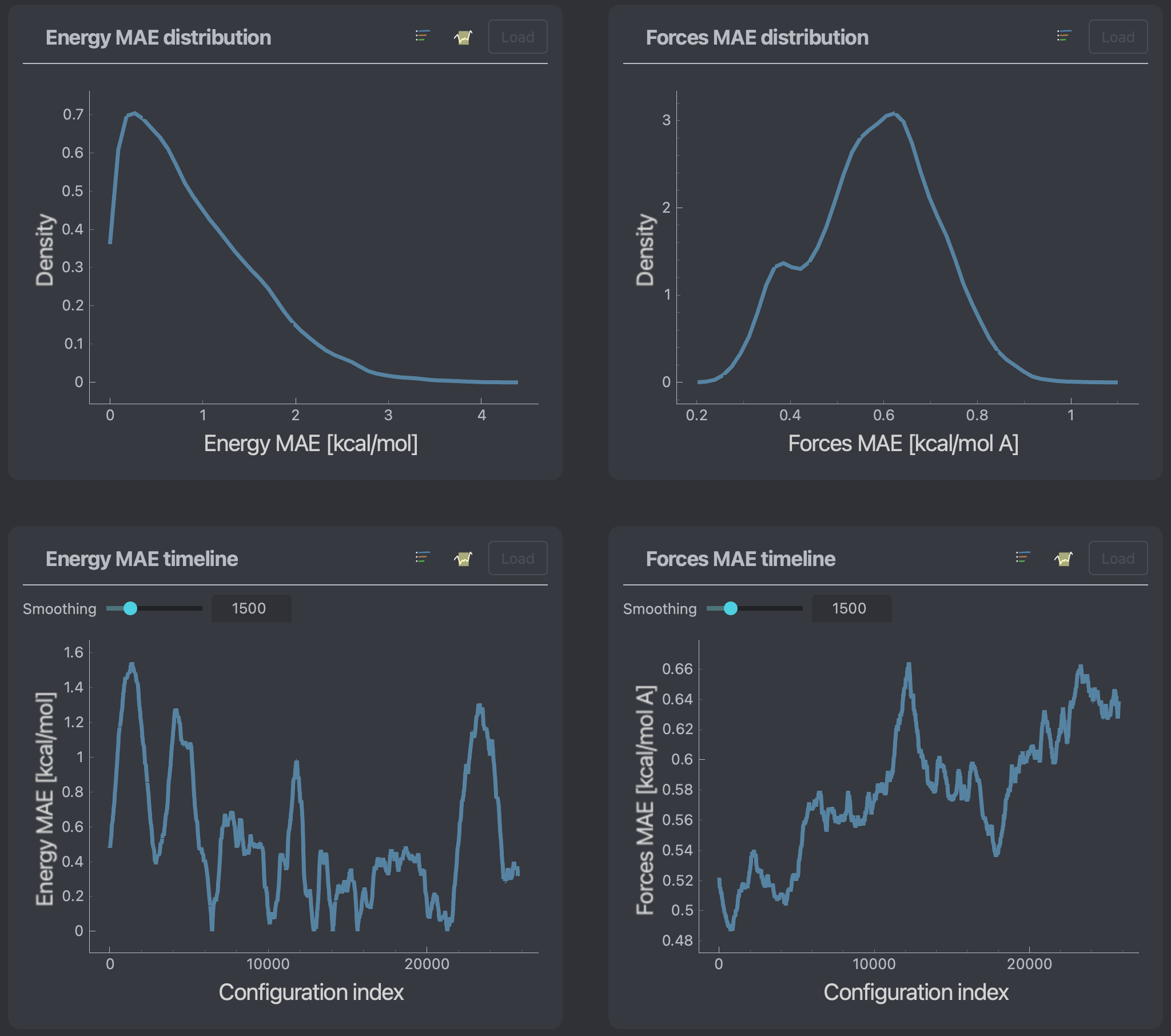}
    \caption{Basic error screen including timelines and distributions for both energy and forces mean average errors across an entire stachyose dataset. In blue, a Nequip model trained on 1000 points of the dataset is used.}
    \label{fig:basicerr}
\end{figure}

\textbf{Prediction error overview:} The first screen to greet the user contains basic information providing an overview of a model's overall performance. This overview includes basic error metrics such as mean average errors (MAEs), root mean squared errors (RMSEs), error distributions, and error timelines throughout a dataset. An example is given in Fig.~\ref{fig:basicerr}. Error distributions are one of the fastest ways to determine a model's performance. From a glance, they provide a rough estimate of the accuracy across a given dataset and what to expect for the low and high end of the error spectrum. Furthermore, one can look at the distribution patterns (such as deviation from a normal curve) to understand if a systematic error occurs on specific data subsets. 

Error timelines show the prediction errors throughout the dataset's lifecycle. This is mostly useful if the data is generated from molecular dynamic simulations, as it allows the user to observe general trends or rough moments in time where an MLFF model's performance might have been lacking. The average window size can be adjusted manually to find a good balance between coarseness and precision.

\bigskip
\textbf{Outlier detection:} Many tools are provided to find outliers in the dataset or in the prediction error profile. While average errors across entire datasets are helpful, they do not provide any indication as to why and where the FF fails. As such, it is essential to determine which configurations cause issues and whether those are outliers in the dataset or important representative geometries. 

Correlation scatters are scatter plots of the predicted values against the true values. While simple, they allow users to visually determine which points, likely far from the expected linear correlation, are worth a second look. Besides, the general shape of a correlation plot gives a good indication of the stability of the model, as in practice, a single bad prediction during a time step can result in the entire dynamics steering into the non-physical territory. 

\begin{figure}[hbt!]
    \centering
    \includegraphics[width=.70\textwidth]{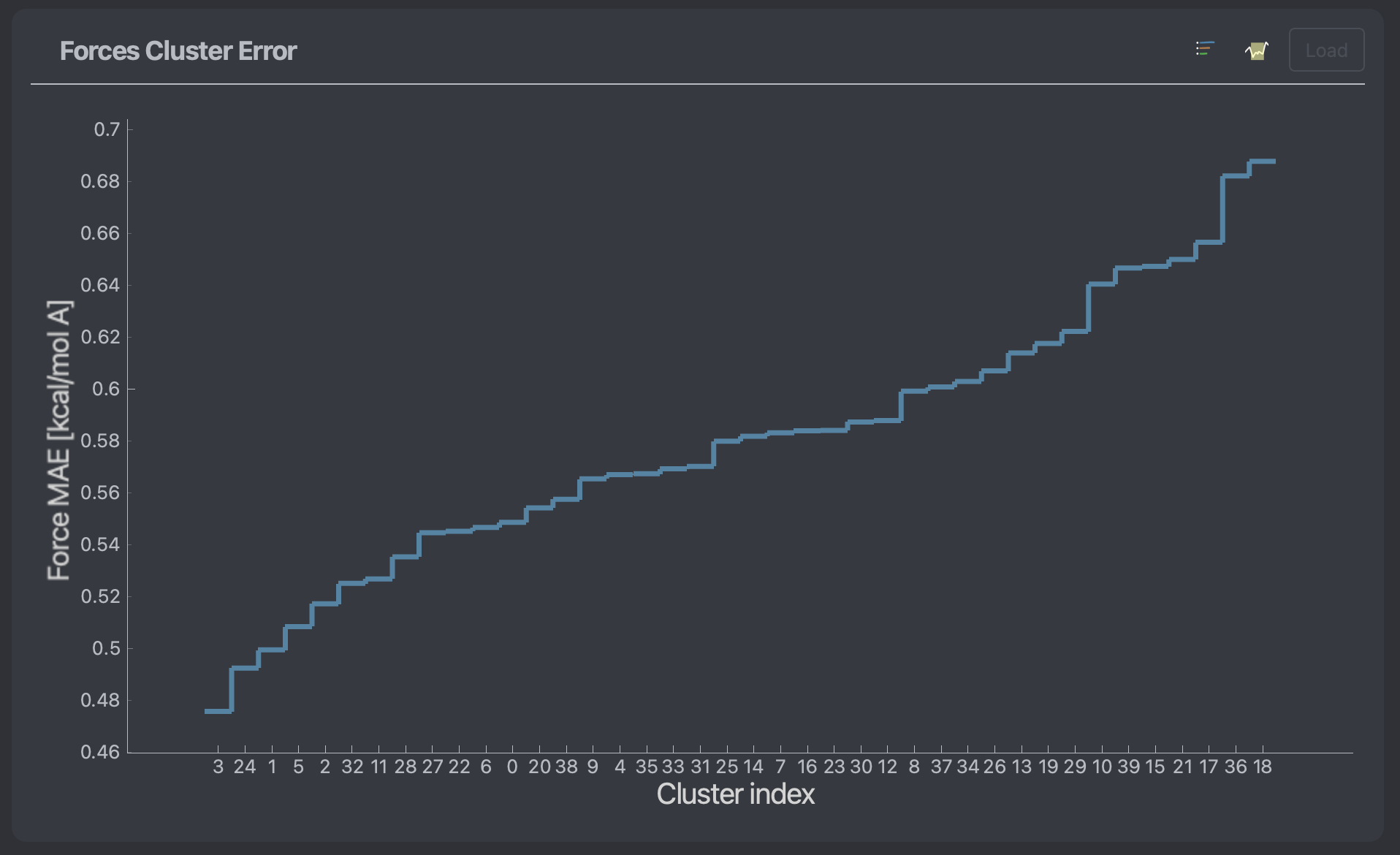}
    \caption{Mean average error of force predictions of a Nequip model trained on 1000 points for 50 different clusters (ordered by ascending error) of a stachyose dataset.}
    \label{fig:clustererr}
\end{figure}

A more in-depth way to visualize a model's stability across areas of configurational space available in a given dataset is through the usage of clustering algorithms~\cite{MLFF}, see example on Fig.~\ref{fig:clustererr}. In essence, the dataset is split into a select number of distinct clusters, each containing a varying amount of configurations. The groups are chosen such as to maximize the similarity between configurations of the same group, thereby splitting the entire datasets into a digestible number of qualitatively different configurations. Error analysis on those different clusters provides an overview akin to an error distribution; however, its discrete nature focuses on types of configuration rather than single points. Thus the user can apply the chemical intuition and knowledge of the molecule at hand to potentially find a) why a model fails on certain clusters but not others, b) what the application range of the model is, or c) outliers in the dataset. 

\begin{figure}[hbt!]
    \centering
    \includegraphics[width=.70\textwidth]{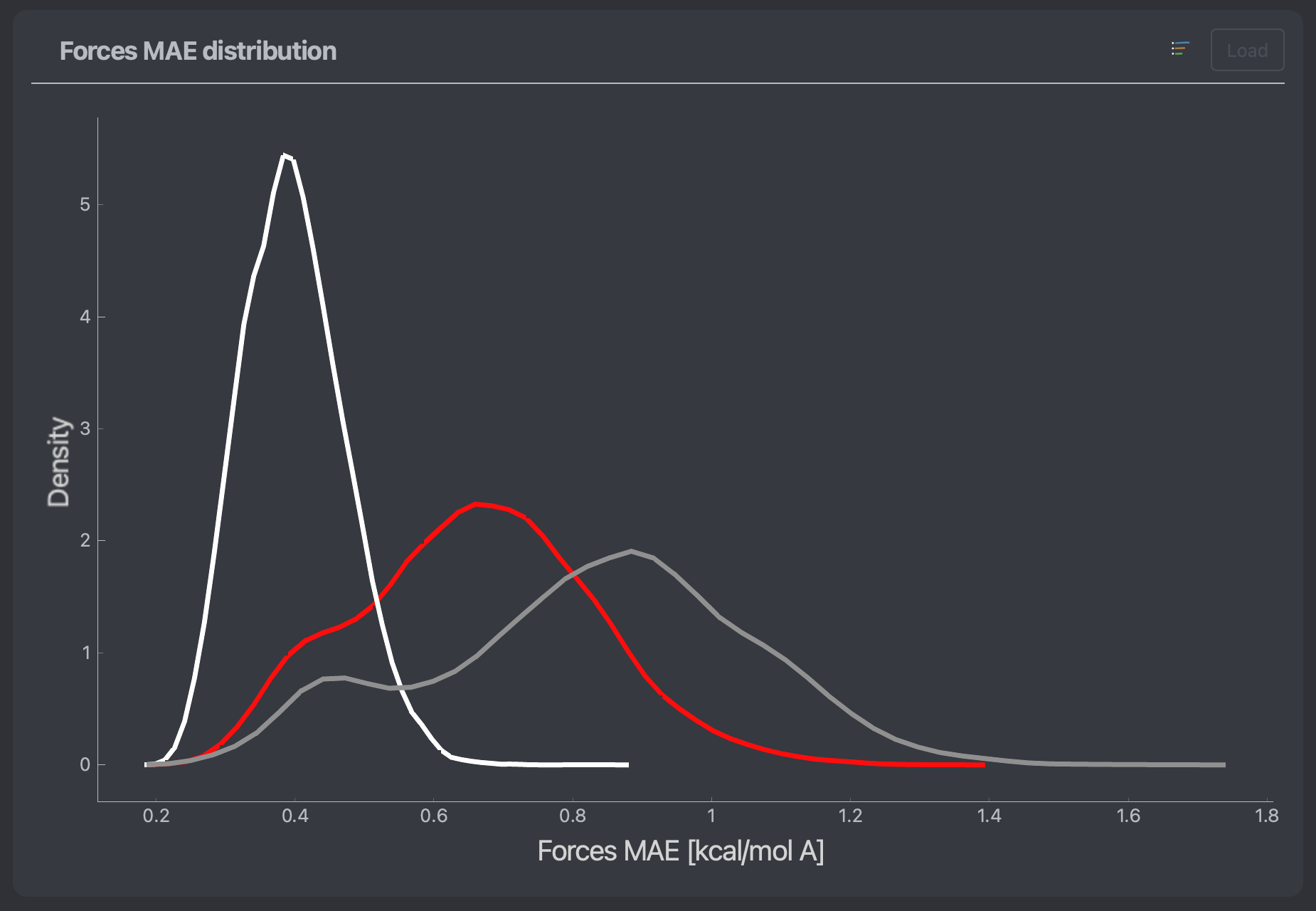}
    \caption{Mean average error density of the force predictions of a Nequip model trained on 1000 points of a stachyose dataset. The densities are done separately for each atom type. Colors correspond to elements (Hydrogen: white, Oxygen: red, Carbon: grey)}
    \label{fig:atomicerr}
\end{figure}

\textbf{Atomic errors:} While the previous plots focus on finding single or groups of configurations with particular prediction error patterns, there is also merit to focusing on models' performance for different atom types. With the increasing chemical composition and structural complexity of the systems of interest for MLFFs, atoms of the same type interact differently with their environment based on their environment composition. Thus, one can expect that ML model predictions vary across atom types. FFAST provides the advantage of visualizing a given model's force prediction error distribution across different chemical elements constituting the system. Alternatively, one can view the error distribution for a selected chemical element across multiple ML models or datasets. An example can be found in Fig.~\ref{fig:atomicerr}

\textbf{3D visualization :} All the previous features offered a way to get a general understanding of a FF performance and find outliers for the models. However, to extract chemical intuition from these results, it is paramount to visualize them in 3D space. FFAST allows most of its plots mentioned above to be used for visualization purposes. Zoom into any subarea of interest and visualize only configurations within that region. Beyond giving form to the points, areas, and clusters of interest, the 3D visualization tool allows for configuration-by-configuration and atom-by-atom basis analysis. One can view the force prediction error for each atom on an individual geometry, or errors averaged throughout all selected configurations. That way, if a specific mechanism is at the root of poor predictions, combining outlier detection (to find the geometries involved) and atomic force prediction error (to see the atoms involved) can provide a path to revealing its cause. 

\begin{figure}[hbt!]
    \centering
    \includegraphics[width=.370\textwidth]{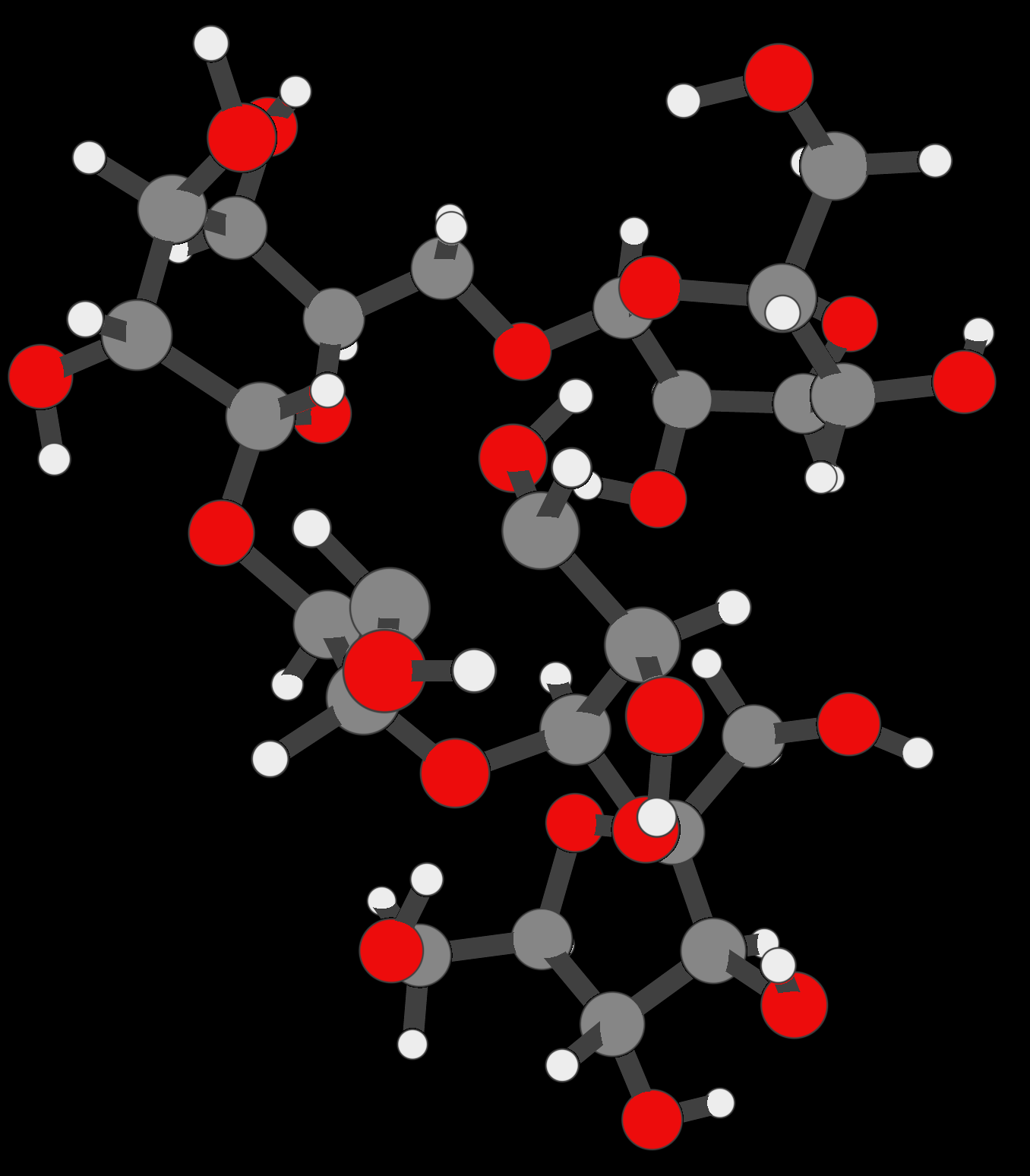}
    \includegraphics[width=.400\textwidth]{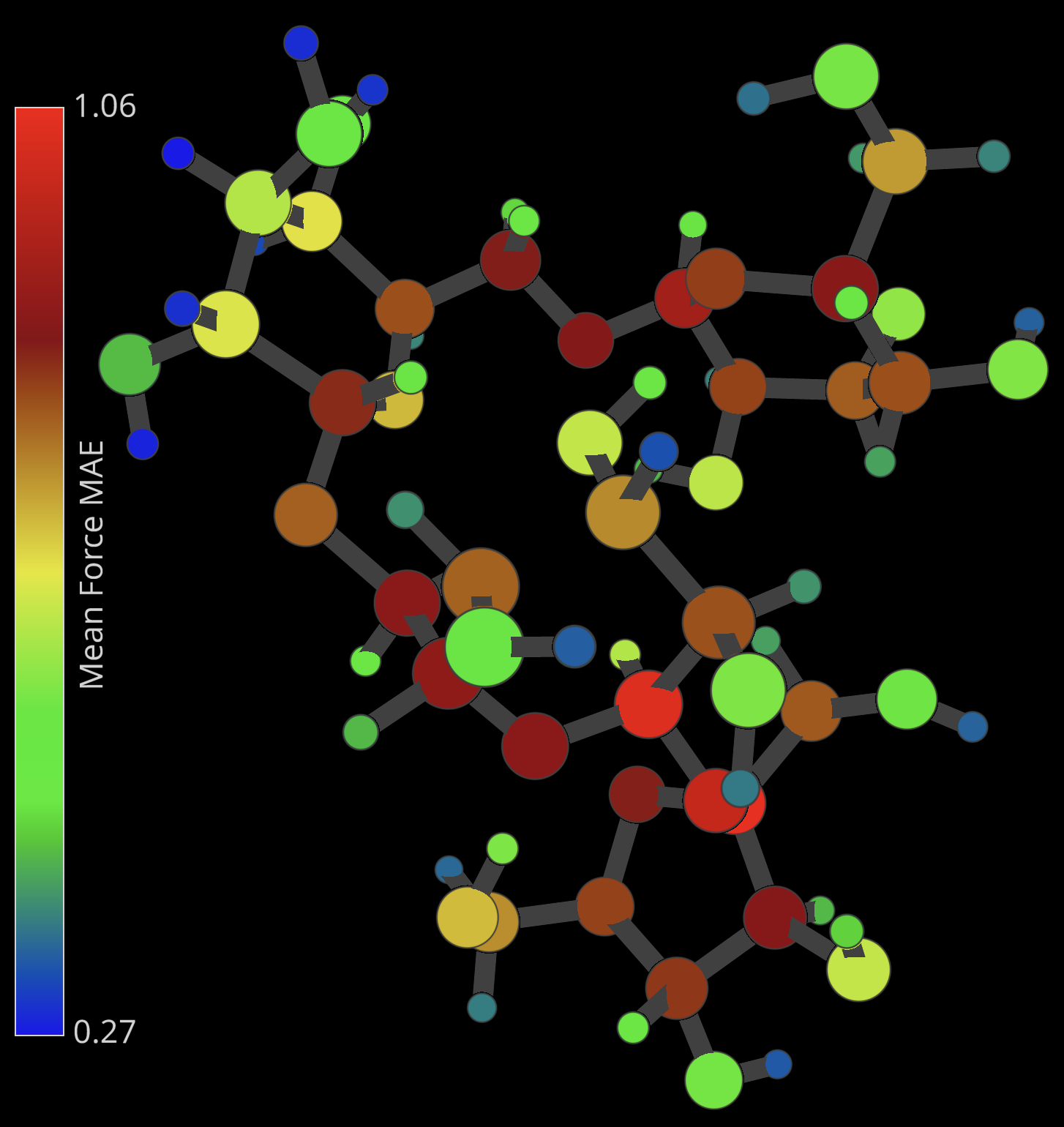}
    \caption{Cutout of the 3D visualiser on the stachyose molecule. (Left) atoms colored according to their element. (Right) atoms colored according to their respective mean average force error as predicted by a Nequip model trained on 1000 points.}
    \label{fig:ferr3d}
\end{figure}

\section{Applications}

After demonstrating the main features of the FFAST, we will apply it to analyze in detail the capabilities of two state-of-the-art ML architectures, namely Nequip and MACE, to reconstruct PESs and FFs of flexible organic molecules.  While both ML models can easily reach below chemical accuracy MAEs on precomputed test datasets, obtaining stable long-time dynamics is a nontrivial task. Therefore, understanding the origin of prediction errors on an atomistic level is crucial for accurate applications and model prediction improvements.

\subsection{Stachyose}
This subsection focuses on the stachyose molecule (C$_{24}$H$_{42}$O$_{21}$), consisting of 87 carbon, oxygen, and hydrogen atoms. The primary ML architecture is the Nequip model trained on the abovementioned dataset and analyzed using the FFAST software. The basic errors screen (see Fig.~\ref{fig:basicerr}) reveals some noteworthy trends. Most of the configurations in the dataset fall within chemical accuracy for the chosen model, as indicated by the energy MAE distribution's main contributor lying below 1~$kcal/mol$. This also applies to the force MAE distribution with the overall MAE of 0.58~$kcal/(mol\,\angstrom)$. Interestingly, the force error distribution demonstrates a well-defined double peak shape, suggesting non-equal force reconstruction for different system components. Also, the forces and energy MAE timelines
shows oscillating behavior clearly distinguishing sets of configurations, reflecting that the molecular dynamics that generated the dataset explored qualitatively different molecular structures throughout the simulation instead of oscillating in or around an equilibrium state. 

The atomic error plot in Fig.~\ref{fig:atomicerr} expands on the overall distributions and distinctly shows a fundamental difference between hydrogen and other elements. The overall MAE and RMSE of each atom type can be found in Table~\ref{tbl:atomicerr}. Importantly, the well-predicted hydrogen atoms, making up almost half of the molecule, highly influence the overall prediction errors. This also largely explains the double peak structure found in the overall force error distribution discussed above. Nevertheless, the secondary peaks in the distributions are also visible on a per-atom basis, notably for carbon but also in oxygen. 

\begin{table}[hbt!]
\begin{center}
\begin{tabular}{| c | c c c c c c c c c | c |}
\hline
 & H & C & $\text{C}_b$ & $\text{C}_r$ & $\text{C}_s$ & O & $\text{O}_b$ & $\text{O}_r$ & $\text{O}_s$  & All \\
\hline
MAE  & 0.40 & 0.83 & 0.92 & 0.78 & 0.77 & 0.66 & 0.89 & 0.82 & 0.57 & 0.58 \\ 
RMSE & 0.55 & 1.10 & 1.23 & 1.04 & 1.02 & 0.90 & 1.20 & 1.10 & 0.76 & 0.82 \\
\hline
\end{tabular}
\caption{Overall force MAE and RMSE in $kcal/(mol\,\angstrom)$ for each atom type in the Stachyose dataset, as predicted by a Nequip model of 1000 training points. The subscripts $b$, $r$, and $s$ indicate filtering for only atoms touching a glycosidic bond, inside the rest of the ring, or in a side chain, respectively.}
\label{tbl:atomicerr}
\end{center}
\end{table}

The reason for the double-peak structure of carbon and oxygen force error distributions is revealed when looking at the average force prediction error for every individual atom in the 3D visualizer; see Fig.~\ref{fig:ferr3d} (right). The molecule consists of a main chain of one furanose (5-membered carbohydrate ring with oxygen) and three pyranoses (6-membered carbohydrate ring with oxygen). Upon inspection, one can notice that the carbons involved in or directly touching the glycosidic bonds between the rings have notably worse errors than those elsewhere on the rings (by about $\sim 18 \%$). A similar situation is found for the oxygens inside said bonds, which have worse predictions than those in the rings or, more significantly, than those in side-chain hydroxyl groups (by about $\sim 56 \%$, see Table~\ref{tbl:atomicerr}). Remembering that a model's actual simulation performance is largely defined by its most problematic atoms rather than an overall mean error is important. Hence, it would be wise to consider this particular model's practical MAE and RMSE to be around $\sim$ 0.9 $kcal/(mol\, \AA)$ and $\sim$ 1.2 $kcal/(mol\, \angstrom)$ respectively. Also, based on this analysis, one could consider introducing atom types based on their chemical environment rather than purely by its nuclear charge, akin to methods employed for empirical force fields with chemically diverse systems~\cite{ewigDerivationClassII2001}.

\begin{figure}[hb!]
    \centering
    \includegraphics[width=.70\textwidth]{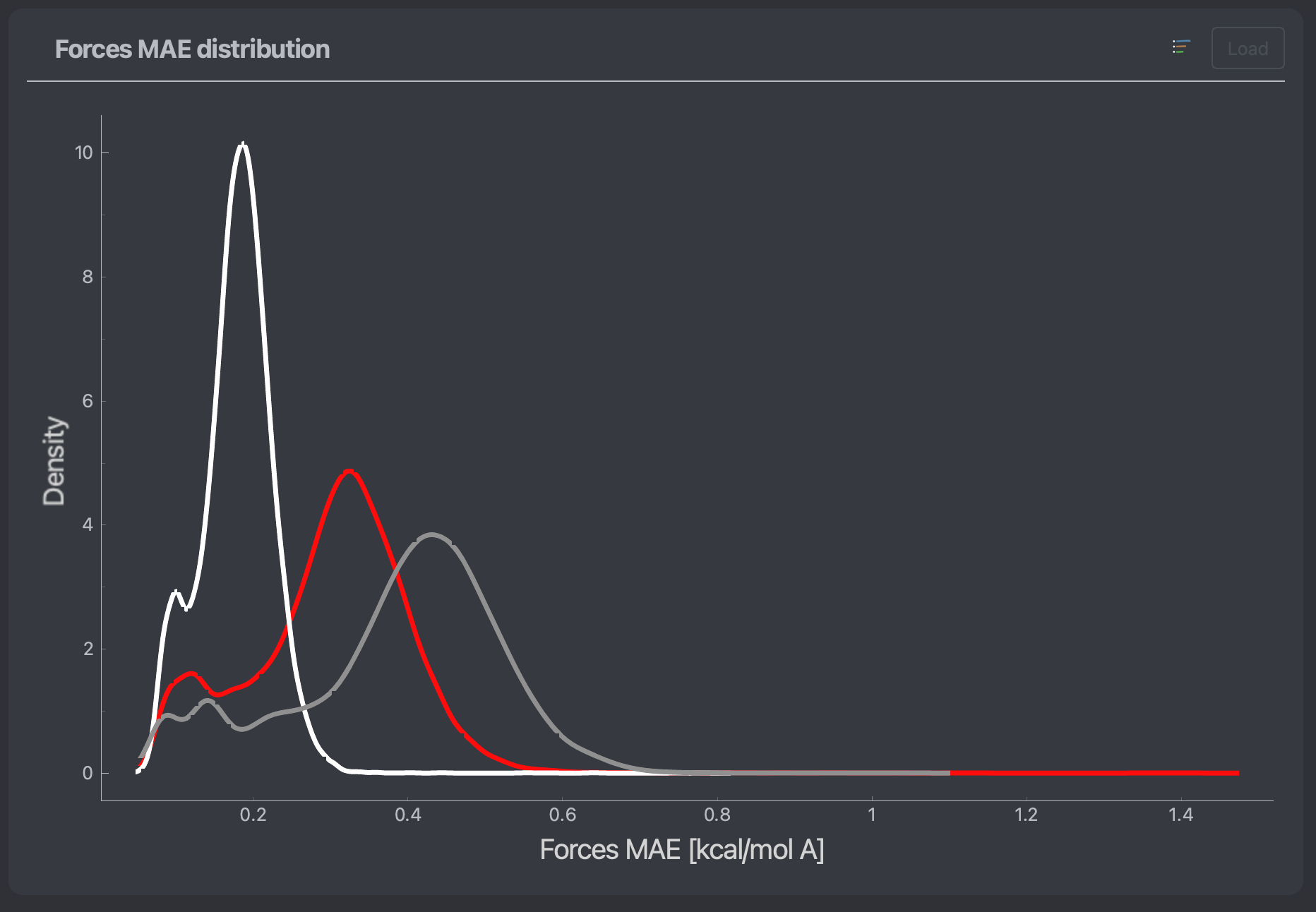}
    \caption{Mean average error density of the force predictions of a MACE model trained on 1000 points of a stachyose dataset. The densities are done separately for each atom type. Colors correspond to elements (Hydrogen: white, Oxygen: red, Carbon: grey).}
    \label{fig:atomicerrstachmace}
\end{figure}
All the prior discussions can be repeated with a separate model, e.g., a MACE model created with the same training set. In Fig.~\ref{fig:atomicerrstachmace}, one can see that a very similar trend can be observed, with even more distinguishable peaks in the error distributions. Also, all the observations made in the 3D visualizer for the MACE model are qualitatively identical to those made for the Nequip model showing significant similarity in both models' performance for the considered system.

\subsection{Docosahexaenoic acid}

\begin{figure}[hbt!]
    \centering
    \includegraphics[width=0.60\textwidth]{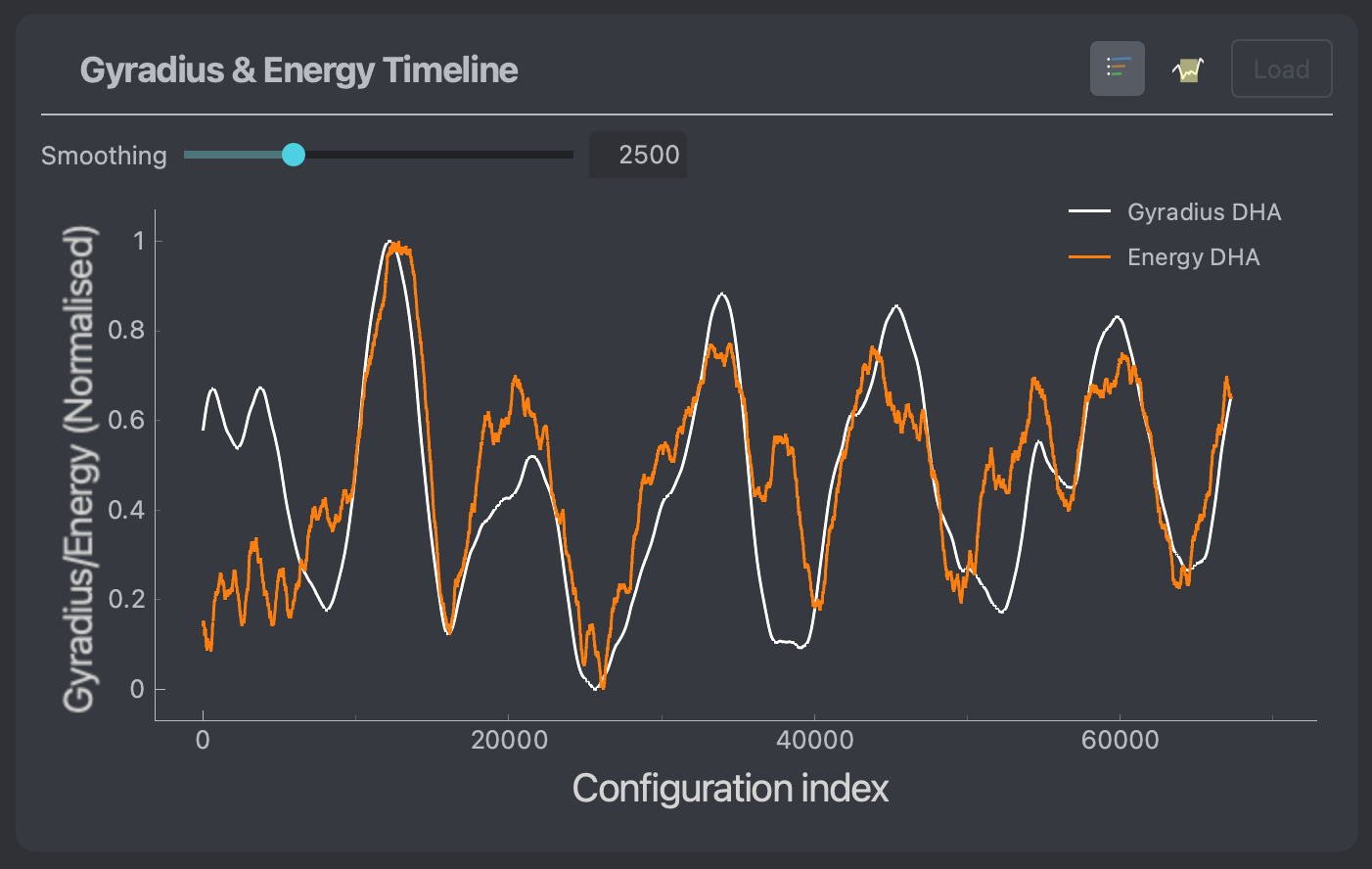}
    \caption{Gyradius (white) and energy (orange) throughout the trajectory of a DHA molecule. Values are averaged over a window of 2500 points to smooth out noise.}
    \label{fig:gyradius}
\end{figure}

\begin{figure}[hbt!]
    \centering
    \includegraphics[width=0.70\textwidth]{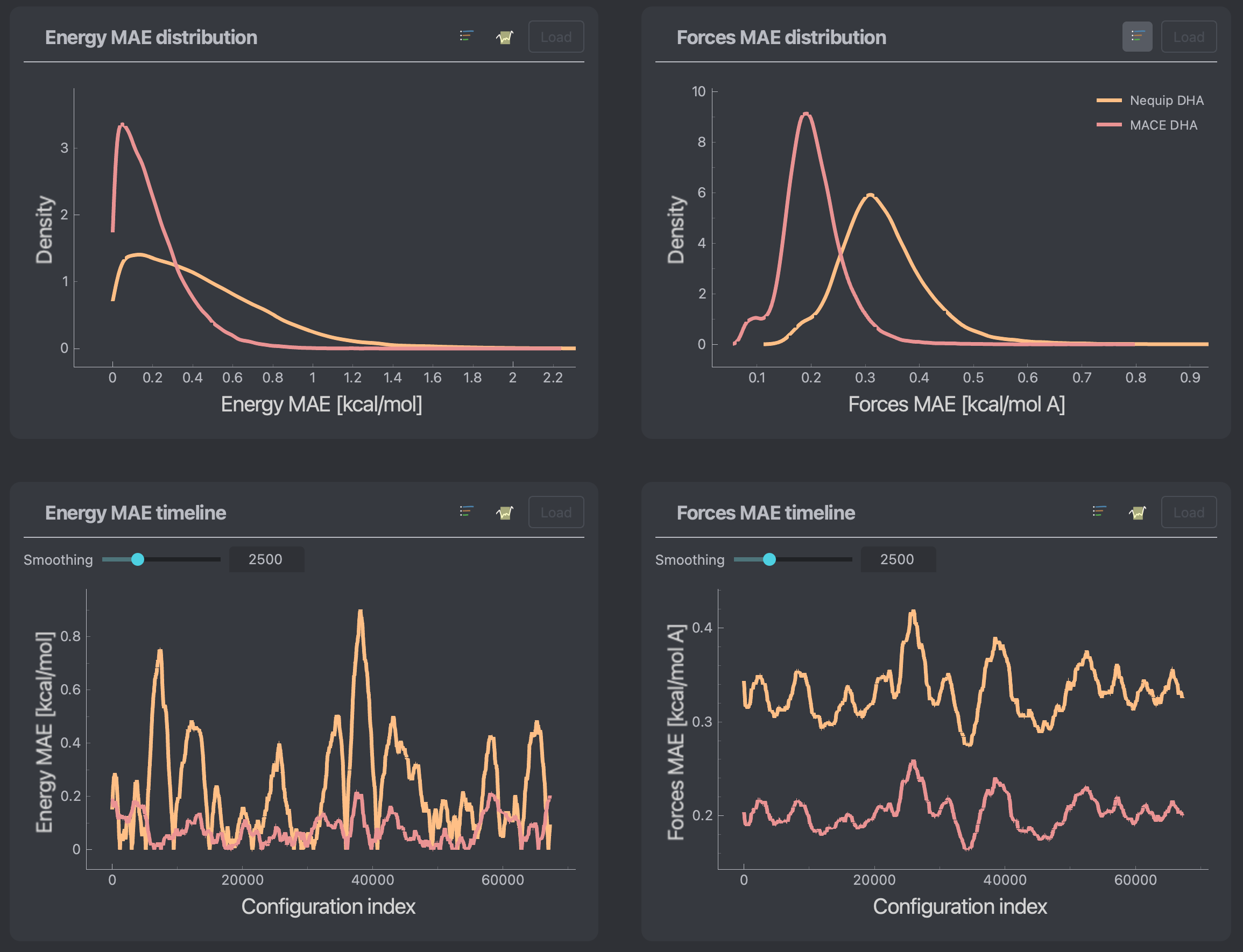}
    \caption{Basic error screen including timelines and distributions for both energy and forces mean average errors across an entire stachyose dataset. In orange and red respectively, Nequip and MACE models trained on 1000 points of the dataset are used.}
    \label{fig:basicerrDHA}
\end{figure}

A second example illustrating the use-case of the FFAST is that of the docosahexaenoic acid (DHA) molecule, a fatty acid with the chemical formula C$_{22}$H$_{32}$O$_2$ for a total of 56 atoms. The molecule comprises a 22-carbon chain with six double bonds and a carboxylic head. DHA is rather flexible due to its long hydrocarbon tail, meaning the molecule can visit various extended and folded states at ambient conditions. Therefore, understanding how well the folding and unfolding processes are represented in the reference dataset is crucial for constructing a reliable MLFF. FFAST provides a simple tool to visualize the folding/unfolding processes by tracking the molecule's radius of gyration (or gyradius), as shown in Fig.~\ref{fig:gyradius}. This does not require any ML model to be loaded. Fig.~\ref{fig:gyradius} shows that, in total, the dataset consists of six compact and six extended states. Moreover, the radius of gyration correlates with the molecule's potential energy after an equilibration period of approximately 10k simulation steps.
One region where the correlation is broken is the fourth folded state (around 40k step). While the molecule is compact, its potential energy is relatively high for those configurations, suggesting the occurrence of some non-trivial interaction patterns demanding further investigation. Interestingly, both ML models trained in this subsection would demonstrate the largest energy MAEs and noticeable force MAEs strictly for this set of structures (see Fig.~\ref{fig:basicerrDHA}), fortifying the previous conclusion.

\begin{figure}[hbt!]
    \centering
    \includegraphics[width=1\textwidth]{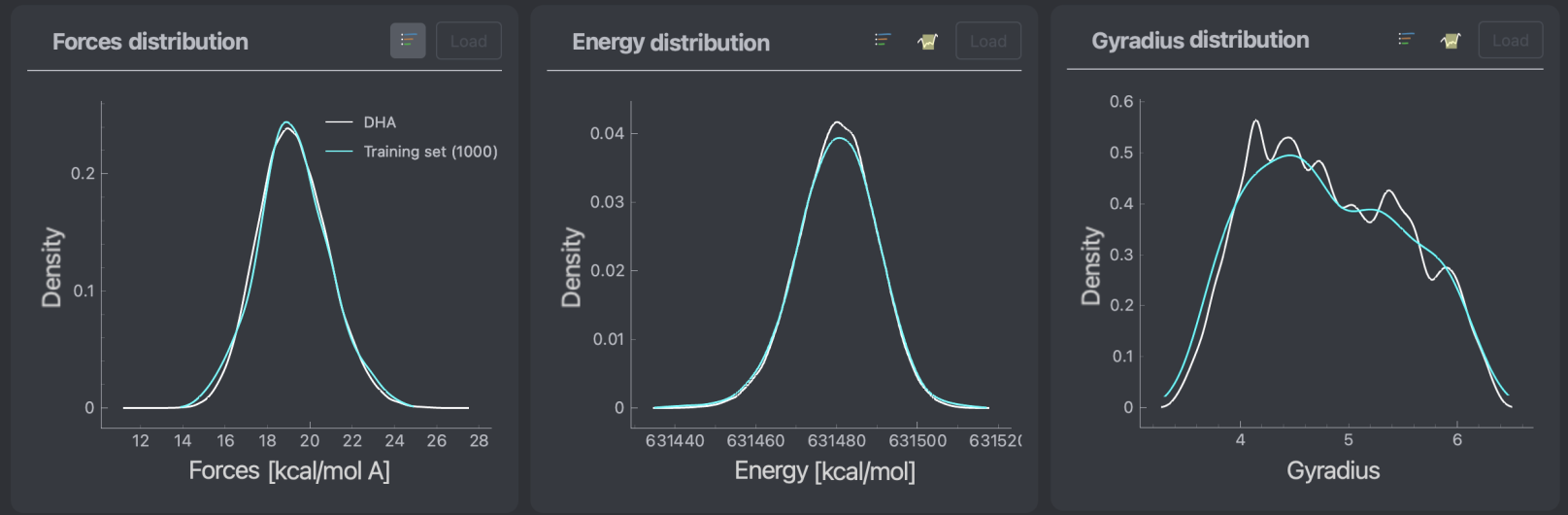}
    \caption{Force distribution (left), energy distribution (middle) and gyradius distribution (right) of the entire DHA dataset (white) against a training set of $1000$ points (cyan). The training set was generated using the sGDML training point selection method.}
    \label{fig:distributions}
\end{figure}

Another essential analysis that can quickly be performed using FFAST and is a prerequisite for training reliable MLFF for flexible molecules: ensuring the quality of the training set. Here, we compare the distributions of gyration radius, forces, and energies within the given (complete) dataset and the generated training (sub)set of $1000$ points, see Figure~\ref{fig:distributions}. The comparison verifies that the training set is indeed representative of the dataset and the obtained ML model should not unexpectedly enter interpolation regimes while within the reference's configurational space. 

With such an insight into the content of the given reference and training datasets, we can build MLFFs aiming at the description of folded and unfolded DHA geometries and the transitions between them. Below, we will compare the performance of two competing state-of-the-art ML architectures under the same conditions. In Fig.~\ref{fig:basicerrDHA}, one can see Nequip (orange) and MACE (red), both generated using the same training and validation set of 1000 points each. One can observe that in this case, the MACE model has better general accuracy than the Nequip equivalent, with each presenting an overall MAE on forces of 0.20~$kcal/(mol\, \AA)$ and 0.33~$kcal/(mol\, \angstrom)$, respectively. Furthermore, the error distribution peaks for the MACE model are narrower for both energy and forces. Interestingly, the difference in accuracy does not come from a better prediction of forces acting on a specific type of atoms within the MACE model. A similar 50~\% decrease in the MAEs is observed for hydrogens, oxygens, and carbons, as can be seen by comparing atomic errors (similar to the analyses shown in Fig.~\ref{fig:atomicerr}). This hints at the overall better performance of the MACE architecture for reconstructing the PES of the DHA molecule.

\begin{figure}[hbt!]
    \centering
    \includegraphics[width=.21\textwidth]{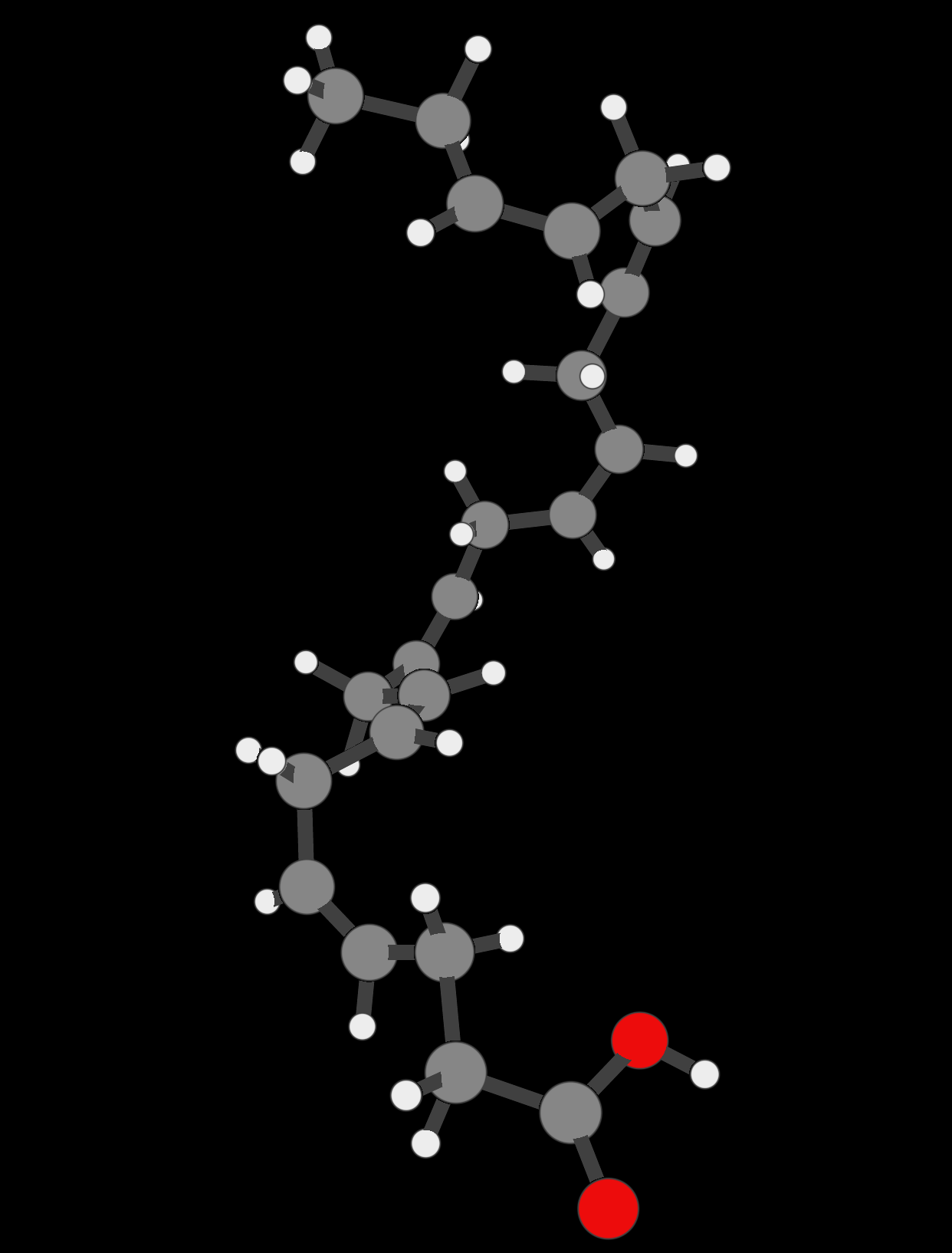}
    \includegraphics[width=.24\textwidth]{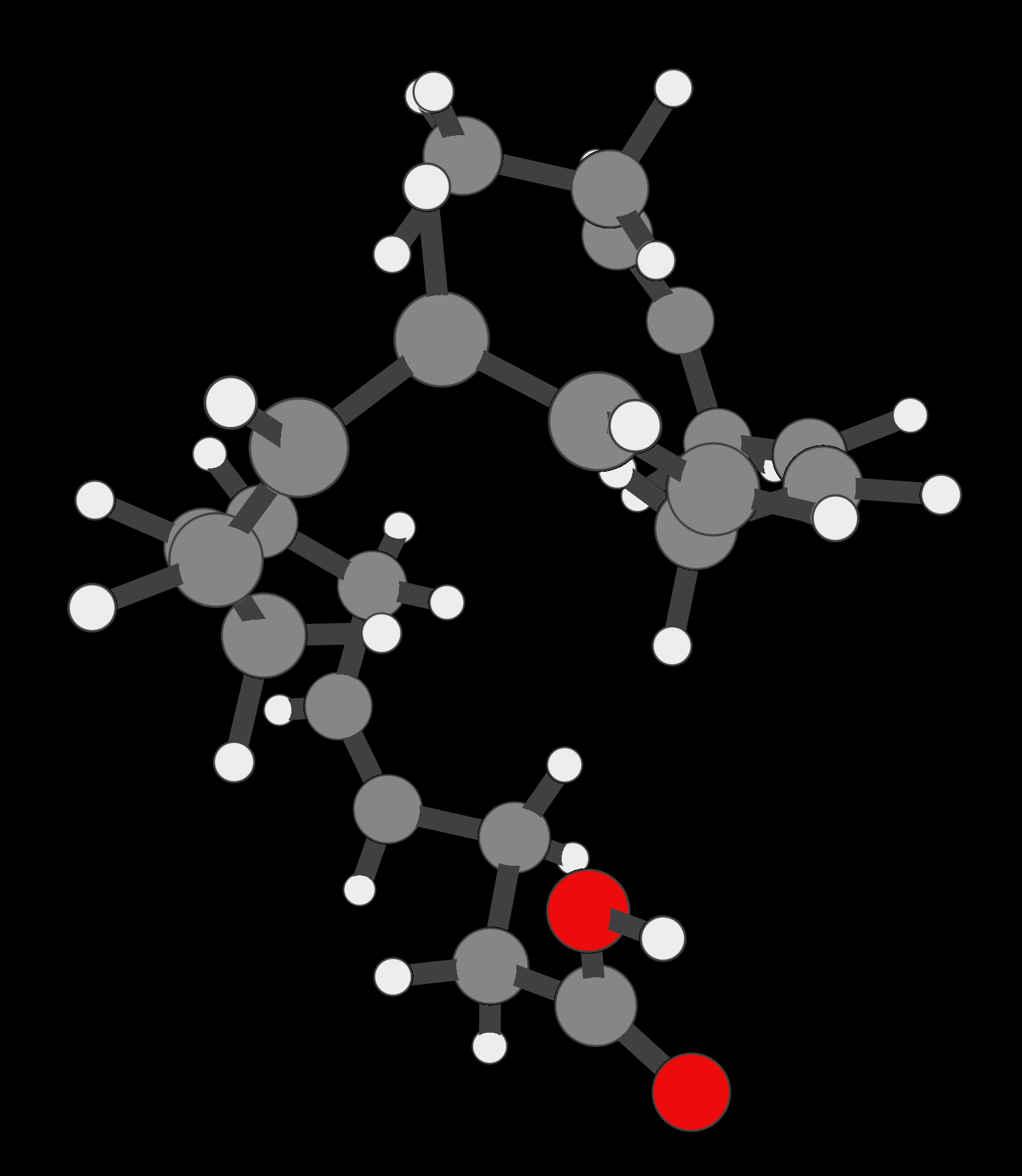}
    \caption{(Left) Example of an extended DHA molecule, as found in the valleys of the MAE timeline and low-error clusters. (Right) Example of a folded DHA molecule, as found in the peaks of the MAE timeline and high-error clusters. The geometries were chosen from the lowest and highest force prediction error cluster, respectively.}
    \label{fig:dhageo}
\end{figure}

Notably, distinct loosely periodic peaks can be observed both in the energy and force MAE timeline. These peaks match between Nequip and MACE, meaning that they are unlikely to be due to an artifact of the models' predictions but rather a fundamental change in geometry throughout those time steps. Zooming in on the valleys/peaks and displaying the configurations in the 3D visualizer reveals that the low error regions correspond to extended geometries while the high error peaks contain folded geometries. An example for each can be seen in Fig.~\ref{fig:dhageo}

\begin{figure}[hbt!]
    \centering
    \includegraphics[width=0.70\textwidth]{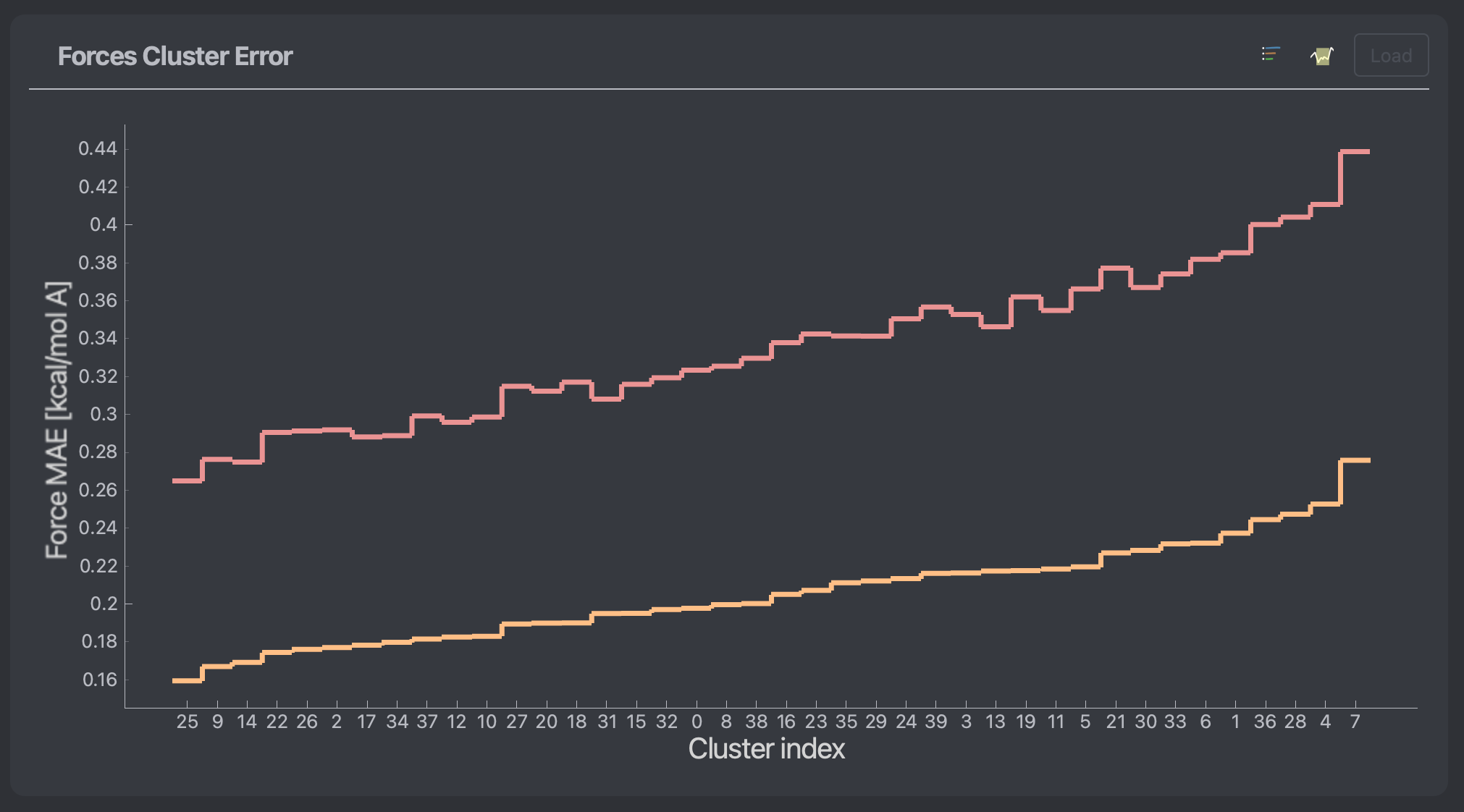}
    \caption{Mean average error of force predictions for 50 different clusters of a stachyose dataset. The models used are a Nequip model (red) and a MACE model (orange) trained on 1000 points. The clusters are ordered in ascending prediction error of the MACE model.}
    \label{fig:dhaclerr}
\end{figure}

The mean force prediction cluster errors (see Fig.~\ref{fig:dhaclerr}) show a relatively flat profile, i.e., the highest average force prediction error on a group of similar geometries is less than two times higher than that of the lowest. Nevertheless, similarly to the error timeline plots, visualizing the high error clusters consistently displays folded configurations and vice-versa. One can find an example of the lowest and highest error clusters in Fig.~\ref{fig:dhageo}.

\begin{figure}[hbt!]
    \centering
    \includegraphics[width=0.48\textwidth]{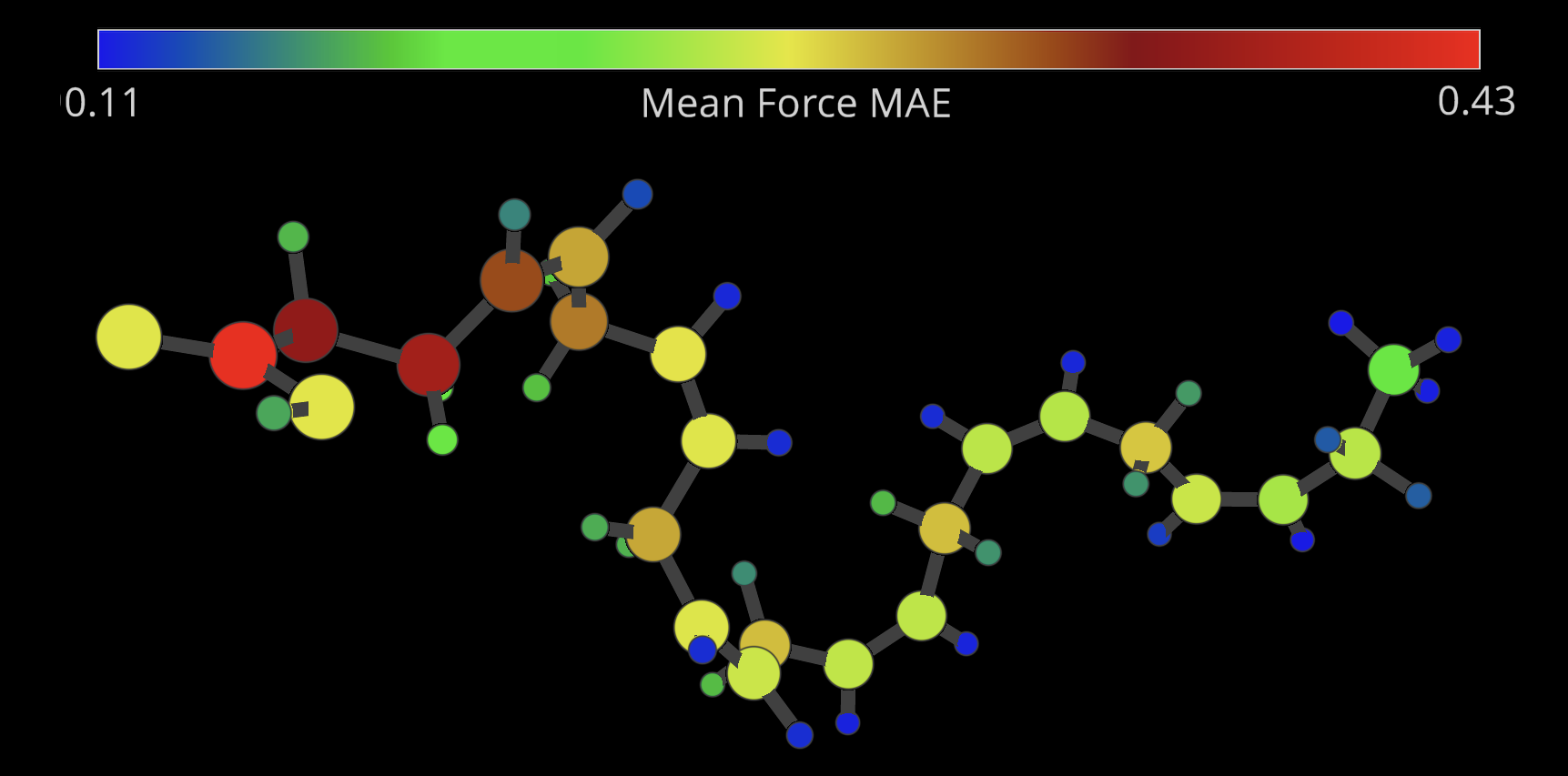}
    \caption{Extended configuration of a DHA molecule. Atoms are colored according to their respective mean average force error as predicted by a MACE model.}
    \label{fig:dhaconf}
\end{figure}

Finally, additional useful information can be obtained by visualizing the average force prediction error on every atom in the 3D visualizer, as seen in Fig.~\ref{fig:dhaconf}. Once again, hydrogens are very well predicted across the board, suggesting that a significant contribution to the low overall prediction error comes from the 32 hydrogen atoms in the molecule. In contrast,  one can see that the carbon atoms inside a carboxylic group and those closest to it (on the left side of the figure) are significantly worse predicted than the rest of the chain. While not shown here, the Nequip model presents a similar overall trend. One can easily explain the observed nonuniform force reconstruction for carbon atoms (both for Nequip and MACE) by combining the information obtained from the FFAST software and our physical/ML intuition. The carbon atoms near the head of the DHA molecule containing oxygen atoms are exposed to a more complex chemical composition of their neighborhood than those in other parts of the molecule. Thus, the descriptor space for such carbon atoms contains more possible states, especially in folded-like configurations. This leads to noticeably more challenging training tasks for the ML models resulting in larger prediction errors. This again hints at the need to reconsider the definition of atomic types based on the atom's environment composition rather than its nuclear weight when we move from simple periodic systems or small molecules to more challenging systems.

All in all, our analysis shows that beyond the insignificant for most practical applications error difference in less than 0.25~$kcal/mol$ for the energy and 0.20~$kcal/(mol\, \angstrom)$ for the forces, both the Nequip and MACE models demonstrate equivalent results for the given test problem. Therefore, one can in principle apply any of these two models to study molecules similar to DHA or stachyose examples.

\section{Outlook}

The applications in this manuscript (as well as many other examples omitted) demonstrate that increased chemical and structural complexity of mid and large-size molecules or periodic systems require a detailed analysis of MLFFs' stability, reliability, and performance. For such cases, standard metrics such as MAE and RMSE are only one piece of the puzzle. High overall accuracies for reconstructing a PES or FF can largely stem from the good performance of an ML model on specific (albeit large) parts of the system; see the case of hydrogens in the examples above. At the same time, forces for other vital system components might be predicted at a significantly lower accuracy. For both the examples of DHA and stachyose, the largest errors for carbon atoms lie in a region where the overall force error distribution shows negligible probability. This effect is clearly observed despite the molecules containing less than a hundred atoms. For more complex systems, such as molecules on surfaces, periodic structures containing rigid crystal frames and molecular interstitials, etc., the difference between mean errors and region-specific errors could become even more prominent. When looking to enhance models and go beyond current capabilities, all these effects can heavily influence the next directions to undertake. To create a reliable way to systematically improve ML models, one first needs to agree on exactly what the current limitations are and which areas need targeted improvements. Furthermore, reaching certain limitations does not necessarily mean the employed ML architectures cannot handle the given tasks. With a detailed insight into the models' performance, one can often significantly increase its applicability by adjusting the training process or the underlying data. Therefore, FFAST or similar software packages should become a staple tool for ML model developers and users.

Another challenge where FFAST can be indispensable is assessing the stability of MLFFs MD trajectories. Providing stable long-time dynamics for complex systems is not at all guaranteed, even for state-of-the-art MLFFs. A long enough simulation eventually reaches a region where extrapolation is necessary, which is likely to generate unphysical forces and heavily influence the rest of the dynamics. These effects can sometimes be easily detected (e.g. breaking of bonds, dissociated hydrogen atoms, etc), but less obvious consequences can also follow. Namely, unphysical behavior can occur without showing apparent damage to the molecular structure within the MD run. Physical/chemical property computed thereafter would be affected without clear evidence to the user that extrapolation regions were reached. While some techniques, such as predicting the model uncertainty, can partially alleviate this problem, the fact that the given ML model is confident in its prediction does not necessarily mean that the prediction is correct. The outlier detection tools and interactive visualization implemented in FFAST provide an efficient alternative instrument to detect such situations without significant human effort. Considering that MLFFs by construction focus first on the most typical interaction patterns, such analyses also highlight non-trivial physics and chemistry that might be present in the system. Therefore, combining FFAST with modern MLFF architectures should also enable a better understanding of the interplay between different types of interactions for large and complex systems. 


\section{Conclusion}
With increased complexities of systems that can be simulated using novel MLFFs comes a need to provide insightful analysis for these models. Even the most advanced MLFFs demonstrate highly heterogeneous predictions across configurational space (CS), invisible to overall error metrics. Thus, the detailed breakdowns of the ML models' performance on different domains of CS and various parts of the system under study should become standard practice. FFAST is an actively developing tool allowing experts and non-experts to get an in-depth insight into the performance of MLFFs or any other force field of choice on any dataset. In this paper, a list of provided features was given and explained, including the example of a best practice scenario. The features were showcased using two datasets of medium-sized flexible organic molecules: docosahexaenoic acid and stachyose. 

Two state-of-the-art ML models (Nequip and MACE) were trained for each dataset. With those, a general FFAST workflow was showcased and the following conclusions followed. Hydrogens are generally significantly better predicted for both molecules than other atoms, considerably lowering the overall ML models' MAEs and RMSEs in hydrogen-rich systems. In contrast, forces acting on carbons and oxygens are unevenly reconstructed. Atoms with chemically or structurally more diverse environments present significantly higher prediction errors compared to the other atoms of the same type. For instance, the prediction errors on DHA rise as the molecule folds, and the primary source of the errors comes from the carboxylic group. 

Beyond a comprehensive analysis of the stability and reliability of ML models, FFAST provides the means to analyze reference data. For instance, one can easily find the number of folding/unfolding processes in a given MD trajectory and the correlation of that with the molecule's potential energy. Another possible application is comparing energy/force distributions in different datasets (training, validation, and test), which should precede any ML model training. As FFAST offers detailed reference datasets and MLFF error analysis, field-specialized knowledge can be applied to enhance MLFFs after identifying the remaining challenges. The software is provided as an open-source project on \url{github.com/fonsecag/FFAST} and is designed to be modulable and expandable to adapt to the user's needs.

\begin{acknowledgments}
We acknowledge financial support from the Luxembourg National Research (FNR) under the AFR project 14593813, FNR DTU-PRIDE MASSENA, and the European Research Council (ERC-AdG grant FITMOL).
\end{acknowledgments}

\section{Bibliography}

\bibliography{FFAST}

\end{document}